\documentclass[journal,draftcls,onecolumn,12pt,twoside]{IEEEtranTCOM}

\usepackage{algorithmicx}
\usepackage{algpseudocode}
\usepackage{cite}
\usepackage{graphicx}
\usepackage{amsmath}
\usepackage{bm}
\usepackage{hyperref}
\usepackage[utf8]{inputenc}
\usepackage[english]{babel}
\usepackage{subfigure}
\usepackage{amsfonts}
\usepackage{epstopdf}
\usepackage{ntheorem-hyper}
\usepackage{algorithm}
\usepackage{algpseudocode}
\usepackage{setspace}
\usepackage{mathrsfs}
\usepackage{caption}
\usepackage{dsfont}
\usepackage[usenames, dvipsnames]{color}
\usepackage{bbm}

\normalsize

\ifCLASSINFOpdf
\else
\fi

\theoremseparator{:}
\newtheorem{theorem}{Theorem}

\newtheorem{case}{Case}
\newtheorem{definition}{Definition}
\newtheorem{lemma}{Lemma}
\begin{document}

\setlength{\baselineskip}{21pt}

\captionsetup[figure]{name={Fig.}}

\title{Adaptive Power and Rate Control for Real-time Status Updating over Fading Channels}

\author{Yalei~Wang,
        Wei~Chen,~\IEEEmembership{Senior Member,~IEEE,}
        \thanks{The authors are with the Department of Electronic Engineering and Beijing National Research Center for Information Science and Technology, Tsinghua University, Beijing, 100084, China, e-mail: yl-wang17@mails.tsinghua.edu.cn; wchen@tsinghua.edu.cn.}

		\thanks{This research was supported in part by the Beijing Natural Science Foundation under Grant No. 4191001, the National Natural Science Foundation of China under Grant No. 61971264 and 61671269, and the National Program for Special Support for Eminent Professionals of China (10,000-Talent Program).}
}

\markboth{IEEE Transactions on Wireless Communications}%
{Submitted paper}

\maketitle
\vspace{-25mm}
\begin{abstract}
	Age of Information (AoI) has attracted much attention recently due to its capability of characterizing the freshness of information. To improve information freshness over fading channels, efficient scheduling methods are highly desired for wireless transmissions. However, due to the channel instability and arrival randomness, optimizing AoI is very challenging. In this paper, we are interested in the AoI-optimal transmissions with rate-adaptive transmission schemes in a buffer-aware system. More specifically, we utilize a probabilistic scheduling method to minimize the AoI while satisfying an average power constraint. By characterizing the probabilistic scheduling policy with a Constrained Markov Decision Process (CMDP), we formulate a Linear Programming (LP) problem. Further, a low complexity algorithm is presented to obtain the optimal scheduling policy, which is proved to belong to a set of semi-threshold-based policies. Numerical results verify the reduction in computational complexity and the optimality of semi-threshold-based policy, which indicates that we can achieve well real-time service with a low calculating complexity.
\end{abstract}

\vspace{-10mm}
\begin{IEEEkeywords}
	\vspace{-5mm}
	Age of information, cross-layer design, AoI-power tradeoff, rate adaptive modulation, constrained Markov decision process,linear programming, probabilistic scheduling, controllable queueing system.
\end{IEEEkeywords}

\IEEEpeerreviewmaketitle
\section{Introduction}
With the development of communication techniques in both present 5G and future 6G\cite{andrews2014will, 6G_Magazine}, communication systems are becoming more and more diverse.
More diverse systems create more diverse requirements on communications. 
In systems that requires real-time transmission, e.g., vehicle-to-vehicle systems and unmanned aerial vehicle systems, there is an increasing interest and demand for monitoring real-time status.
In such systems, optimize the end-to-end latency can sometimes be trivial.
For example, if the source update its status once an hour and transmit the updated information within seconds, the end-to-end latency is at most one minute. 
However, as the source update its status once an hour, the freshness of its status at the receiver would be no less than half an hour.
That is, the real-time property of this system is very poor. 
Therefore, information freshness is often considered as an indispensable parameter as important as latency in these systems.

As information freshness is different from the traditional Quality of Service (QoS) guarantees like end-to-end latency and throughput, a new metric, namely age of information, has been widely adopted to characterize information freshness \cite{2012Age_Definition}. 
By labeling each packet with a time-stamp of its born time, the age of each packet can be marked. AoI is defined as the expectation of the age of the most recently received packet at the receiver. 
As AoI can be affected by updating rate, channel condition, transmission rate and so on, it is very challenging to optimize AoI.
Based on the differences of optimization methods, we classified the previous works in this domain into two categories.

One line of works optimized AoI through adjusting the scheduling strategies \cite{2012Age_Definition, 2012Age_MultipleSource, 2016Age_LCFS, 2016Age_LCFS2,ARQ_AoI, tang2019scheduling,tang2019minimizing,wang2019minimizing,kadota2016minimizing, kadota2018scheduling, talak2018scheduling,talak2018optimizing,kadota2018optimizing,yates2017age,maatouk2020optimality}.
Different from latency, there exists an optimal updating rate for AoI.
In First-Come-First-Served (FCFS) M/M/1 systems, the authors in \cite{2012Age_Definition} gave the analytical expression of the optimal updating rate.
For multiple sources systems, the authors further resolved the optimal updating rate in the presence of interfering traffic in \cite{2012Age_MultipleSource} and cache updating systems in \cite{yates2017age}.
The authors in \cite{maatouk2020optimality} proved the optimality of the Whittle's index policy in multi-user systems.
In multi-user systems with controllable updating process, the authors in \cite{tang2019scheduling} used CMDP to model their scheduling method.
They also showed that the optimal scheduling policy has a threshold structure and extended their work to communication systems with Markov channel model \cite{tang2019minimizing}.
Threshold structure was also proved to be optimal in \cite{wang2019minimizing}, where the authors considered a resource constrained scenario.
The queuing method is also an important factor that affects AoI.
The study in \cite{2016Age_LCFS} and \cite{2016Age_LCFS2} showed that the Last-Come-First-Served (LCFS) principle, as well as re-transmission, can successfully avoid the increments of peak AoI.
The Hybrid Automatic Repeat reQuest (HARQ) and re-transmission protocols were taken into account in \cite{ARQ_AoI}, where preemptive scheduling policies were presented to optimize AoI.
The authors in \cite{kadota2018scheduling,kadota2016minimizing,kadota2018optimizing} showed that a greedy scheduling policy can reach the optimal AoI in symmetric networks.
For general networks, the authors developed three different scheduling policies to optimize AoI.
In systems with time varying channels, the authors in \cite{talak2018scheduling,talak2018optimizing} optimized peak AoI with scheduling policies based on virtual queue. Moreover, they proposed a sub-optimal scheduling policy based on age.

The other line of works optimized AoI through energy allocation and cross-layer control \cite{2014PacketLoss, 2015Age_Energy, 2015Age_Power, sun2017update, arafa2018age1,yang2017optimal, borkar2017opportunistic,wang2017delay,wang2018joint,hsu2017age,jiang2018decentralized,lu2018age,franco2016lupmac}.
When the source node can manage the arriving data and the transmission, discarding packets once the source node is busy could improve the average AoI and the peak AoI \cite{2014PacketLoss}.
The authors in \cite{sun2017update,2015Age_Energy} studied the optimal scheduling policy, in which the power function is considered.
For wireless transmissions, the optimization of AoI is investigated in \cite{hsu2017age,jiang2018decentralized,lu2018age}.
The authors presented an optimal threshold policy to achieve a better AoI.
In \cite{2015Age_Power,arafa2018age1,yang2017optimal}, the optimization problem of AoI in energy harvesting systems were considered.
For multi-user systems with finite queue length, AoI is optimized through considering an energy efficient method in \cite{borkar2017opportunistic}.
In our precious work, cross-layer scheduling method has been a powerful approach to minimize the latency in communication systems \cite{wang2017delay,wang2018joint,xiaoyu_TCOM,shaoling_infocom,liujuan_journal}.
Based on the similarity of latency and AoI, cross-layer scheduling can also be an important approach to minimize AoI. 
Actually in some previous work \cite{franco2016lupmac,tang2019scheduling}, the authors have already used a cross-layer structure to optimize AoI.
In addition to the traditional AoI definition, there are also some other definitions of information freshness.
In \cite{random_process}, a general cost function of estimation errors was presented to measure the difference between the received signal and the original signal.
This cost function consisted of both transmission consumption and information freshness.
For counting processes \cite{counting_process, wang2019realtime}, a similar cost function is present to measure the reconstruction of the original signal as well as the freshness of information.

To the best of our knowledge, although AoI have been optimized in various methods, the design to optimize AoI that combines time-varying wireless channel with average power constraint in a buffer-aware system has not been studied.
To fill this gap, we aim at the optimal tradeoff between AoI and average power consumption in wireless transmissions over time-varying fading channels.
To ensure the integrity of the information from the source, a buffer is equipped to store the untransmitted data.
When the transmission happens, the transmitter sends packets in the buffer to the receiver with an adaptive transmission rate, which is based on the channel state and the buffer state.
Inspired by the optimization of latency, we propose a probabilistic scheduling method imposed on the real-time channel state, the age of packets in the buffer, and the age of the receiver to character the update process of AoI.
Based on this probabilistic scheduling policy, we formulate the update process of AoI as a CMDP.
Then, we give the mathematical expression of AoI and average power consumption and present an optimization problem to minimize AoI.
However, the variable space of this optimization problem is too large, which makes the optimization problem too difficult to solve.
Fortunately, we find is possible to convert the optimization problem into an LP problem through linear transformation.
To solve the LP problem, we show that the optimal policy can be found by only searching within the semi-threshold-based policies, whose LP problem is much easier to solve.
To further reduce computational complexity, an algorithm is presented to obtain the optimal scheduling policy.
Moreover, numerical results show that the optimal scheduling policy has a threshold-based structure imposed on the channel state and the real-time AoI at both ends of the transceiver.

We summarize the main contributions of this paper as follows:
\begin{itemize}
	\item We propose a cross-layer model to characterize the evolution of AoI. To promise the data integrity of the source, we adapt a buffer at the source and do not drop packets, which makes the optimization of AoI more challenging.
	\item To guarantee the optimality of the scheduling, the age of both the source and the receiver are considered. Through variable substitution, we formulate the original problem as an LP problem and obtain the optimal AoI-power tradeoff.
	\item By proving that the optimal scheduling policy exists in semi-threshold scheduling policy, we reduce the complexity of searching the optimal scheduling policy from exponential to polynomial.
\end{itemize} 

This paper is organized as follows. We introduce our system model in Section \ref{system_model_sec}. In Section \ref{problem_formulation}, we present a probabilistic scheduling policy and formulate our system model by CMDP. In \ref{analysis}, we first give the expression of AoI and the average power consumption, based on which an linear programming problem is formulated. Then we present an algorithm to obtain the optimal scheduling policy. Numerical results are present in Section \ref{numerical_results}. Finally, we draw the conclusion in Section \ref{conclusion}.


\section{System Model}\label{system_model_sec}
	\begin{figure}[tbp]
		\centering
		\includegraphics[width = 1\textwidth]{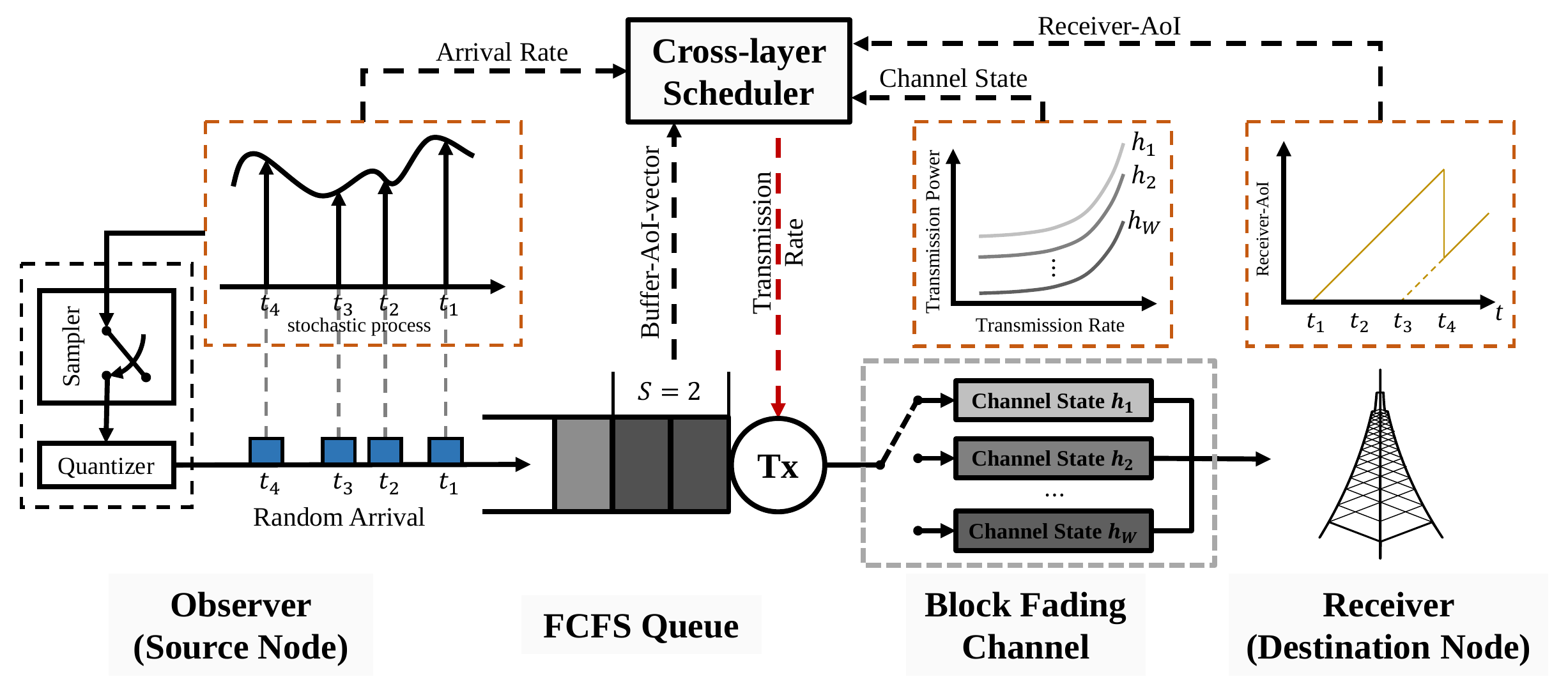}
		\caption{System Model}\label{system_model}
	\end{figure}
	In this paper, we focus on a wireless communication system where the source node transmits real-time data packets to the destination through an FCFS queue.
	As shown in Fig. \ref{system_model}, the time is slotted into intervals with the same length.
	At the beginning of each time slot, the data packets sampled from a stochastic process arrive at the buffer.
	Let us denote the arrival rate of the data packets by $\lambda$.
	In each time slot, we assume that at most one data packet can be generated.
	We denote the number of arrivals at the $n$th time slot by $a[n]$.
	Then we have $a[n]\in\{0,1\}$ and 
	\begin{equation}
		\mbox{Pr}\left\{a[n]=a\right\}=
		\left\{
		\begin{aligned}
		&1-\lambda,&&a=0,\\
		&\lambda,&&a=1.
		\end{aligned}
		\right.
	\end{equation}
	
	The data packets is temporarily stored in the buffer and transmitted on FCFS basis.
	That is, the oldest packet in the buffer is transmitted with the highest priority.
	We assume that the capacity of the buffer is sufficiently large so that the buffer overflow is ignored.
	In each time slot $n$, we denote the number of packets in the buffer by $q[n]$ and the number of transmitted packets by $s[n]$.
	Then we have
	\begin{equation}
		q[n+1] = \max \{q[n]-s[n]+a[n+1],0\}.
	\end{equation}

%
%
%
%
%

	In our system, the packets are transmitted over a block fading channel.
	We assume that the channel status remains stable in one time slot and follows an independently identically distribution (\textit{i.i.d.}) process across different time slots.
	We denote by $h[n]$ the channel coefficient of the $n$th time slot.
	With a given channel state $h$ and a transmission rate $s$, we define the power consumption by function $P(h,s)$.
	For general communication scenarios, a greater transmission rate corresponds to a higher power cost and a lower power efficiency \cite{el2002energy}.
	Therefore, we assume that the power consumption function $P(h,s)$ is monotonically increasing and convex in $s$ for a fixed $h$ in our system.

	We further quantify the fading channel into a W-state channel based on the modulus of the channel coefficient.
	More specifically, we quantify the channel coefficient into $W$ states by $\{h_1,h_2,\cdots,h_W\}$, which satisfies $\delta=h_0<h_1<\cdots<h_W=\infty$.
	In each time slot, if the channel gain ranges in interval $[h_{\omega-1},h_{\omega})$, we define that the channel is at $channel~state~\omega$.
	A smaller channel gain represents a worse channel condition, which implies that channel state 1 and channel state W represent the worst and best channel condition, respectively.
	At the beginning of each time slot, the channel state is reported via a Channel State Information (CSI) channel.
	Let us denote by $c[n]$ the channel state at the $n$th time slot.
	Based on the probability density function of the channel coefficient, we define the probability distribution of different channel states as
	\begin{equation}
		\mbox{Pr}\left\{c[n]=\omega\right\} = \alpha_\omega,~1\leq\omega\leq W,
	\end{equation}
	where $\alpha_\omega\in[0,1]$ and $\sum\limits_{\omega=1}^W\alpha_\omega=1.$

	As the throughput of the transmitter is limited, the number of packets that can be transmitted in one time slot is upper bounded by $S$.
	For each time slot $n$, we have $s[n]\in\{0,1,\cdots,S\}$.
	If the channel is at channel state $\omega$, we denote the transmission power when $s$ packets are transmitted by $P_{\omega,s}$.
	Based on Shannon–Hartley theorem, successful transmission requires too much power when the channel condition is bad.
	To avoid excessive power consumption, we silence the transmitter when the channel gain is smaller than a positive number $\delta$.
	That is, the transmission rate $s=0$ when $h$ ranges in $[0,\delta)$.
	For every channel state $\omega$, if the transmission rate $s=0$, no power would be consumed. 
	Thus we have $P_{\omega,0}=0$ for $1\leq\omega\leq W$.
	Basically, for the same channel state $\omega$, more transmission power is required when more packets are transmitted, which leads to $P_{\omega,1}<P_{\omega,2}<\cdots<P_{\omega,S}$.
	Similarly, for the same transmission rate $s$, more power is required when the channel state is bad, which leads to $P_{1,s}>P_{2,s}>\cdots>P_{W,s}$.
	Moreover, due to the convexity of $P(h,s)$ in $s$, we have $\frac{P_{\omega,s_1}}{/s_1}<\frac{P_{\omega,s_2}}{s_2}$ when $s_1<s_2$.
	In our system, the transmission power supports sufficient high signal-to-noise ratio so that the transmission failures can be ignored.
	
	In the $n$th time slot, we denote by $b[n]$ the born time of the most recently received packet after the arrival process and before the transmission process.
	Since the transmission of time slot $n$ has not happened yet, we have $1\leq b[n]\leq n-1$.
	For each packet in the buffer, as the throughput of the transmitter is upper bounded by $S$, we know that only the oldest $S$ packets have the opportunity to be served.
	We define the age of a packet as the interval from its birth to the present time slot.
	Let us denote by $b_s$ the born time of the $s$th oldest packet in the buffer.
	If there are $k$ packets, which is less than $S$, in the buffer and $k<S$, we set $b_s[n] = n+1$, where $k+1\leq s\leq S$, to denote that packets might arrive in the future.
	Based on the born time of the packets, we give the mathematical expression of the age of the packets in the buffer, which we call \textit{buffer-AoI-vector}, as follows.
	\begin{definition}
		The buffer-AoI-vector $\bm{A}_b[n]$ at the $n$th time slot is defined as a row vector, in which all the age of the packets in the buffer is listed.
		\begin{equation}
		\bm{A}_b[n] = (A_S[n],\cdots, A_2[n], A_1[n]).
		\end{equation}
		The $k$th component of vector $\bm{A}_b[n]$ is given by
		\begin{equation}\label{rank_seq}
		A_s[n] = n-b_s[n],
		\end{equation} 
		where $ s=1,2,\cdots,S$.
	\end{definition}
	\begin{definition}\label{def_AoI}
		The receiver-AoI $A_r[n]$ at the $n$th time slot is defined as the age of the most recently received packet, which is given by
		\begin{equation}
		A_r[n] = n-b[n].
		\end{equation}
	\end{definition}
	
	From Eq. (\ref{rank_seq}), we know that the age of packets in the buffer satisfies $A_1\geq A_2\geq\cdots\geq A_S$, where the $k$th equal sign holds if and only if there are less than $k$ packets in the buffer.
	Since the packets are served on the FCFS basis, we know that the age of the oldest packet in the buffer is smaller than the receiver-AoI, i.e. $A_1<A_r$.
	As the update of receiver-AoI is closely related to the buffer-AoI-vector and the transmission rate, we combine the buffer-AoI-vector and the receiver-AoI as the \textit{system-AoI-vector} of our system.
	The specific definition of system-AoI-vector is given as follows.
	\begin{definition}\label{system_state_def}
		The system-AoI-vector $\mathcal{T}$ is defined as a row vector, whose components are the combination of the buffer-AoI-vector and the receiver-AoI.
		\begin{align}\label{def_3}
		\mathcal{T}	
		= (A_S,\cdots, A_2,A_1, A_r)
		\end{align}
	\end{definition}

\section{Probabilistic Scheduling Policy}\label{problem_formulation}
	To minimize average receiver-AoI, we characterize the scheduling policy with probabilistic scheduling. Based on the probabilistic scheduling policy, we summarize the transmission process in our system as a CMDP.

		Based on the definition of system-AoI-vector, we present a buffer and channel aware probabilistic scheduling policy.
		In the $n$th time slot, the transmission rate $s[n]$ is determined by the system-AoI-vector $\mathcal{T}[n]$ and the channel state $c[n]$.
		By presenting $s[n]$, $\mathcal{T}[n]$, and $c[n]$ by $s$, $\mathcal{T}$, and $\omega$, we define the probability that $s$ packets are transmitted in one time slot as $f_{\mathcal{T}}^{\omega,s}$, which is given by
		\begin{equation}\label{def_sche}
			f_{\mathcal{T}}^{\omega,s} = \mbox{Pr}\{s[n]=s|\mathcal{T}[n]=\mathcal{T}, c[n]=\omega\}
		\end{equation}
		
		Also, we notice that the transmission rate would not exceed the number of packets in the buffer in one particular time slot.
		In the $n$th time slot, if there are $K$ packets in the buffer, we have
		\begin{equation}\label{scheduling_eq}
			\mbox{Pr}\{s[n]=s\}= \left\{
				\begin{aligned}
					&f_{\mathcal{T}}^{\omega,s}, &&\quad0\leq s\leq \min\{\omega,s\},\\
					&0,&&\quad s>\min\{\omega,s\}.
				\end{aligned}
				\right.
		\end{equation}
		As the transmission rate $s$ satisfies $s\in\{0,1,\cdots,\min\{\omega,s\}\}$, we have
		\begin{equation}
			\sum_{s=0}^{\min\{\omega,s\}}f_{\mathcal{T}}^{\omega,s}=1.
		\end{equation}
		If there are no packets in the buffer, i.e., $K=0$, the transmitter keeps silent under this circumstances.
		
		Noticed that the channel states across different time slots follow an \textit{i.i.d.}, we can reduce one dimensional of the scheduling policy and simplify the scheduling parameters as
		\begin{equation}
			f_{\mathcal{T}}^{s} = \sum\limits_{\omega=1}^{W}\alpha_\omega f_{\mathcal{T}}^{\omega,s}.
		\end{equation} 
		For the convenience of expression, we denote a scheduling policy by a infinite dimensional matrix $\bm{F}$. We denote the field of all policies by $\mathcal{F}$. Then we have $\bm{F}\in\mathcal{F}$.

		After we have given the scheduling policy, we formulate the update process of AoI into a Markov chain, whose Markov states consist of the system-AoI-vector $\mathcal{T}$ and the channel state $c$.
		Specifically, we pick the system-AoI-vector after the arrival process and before the transmission process in one time slot.
		As we mentioned before, the channel states across different time slots follow an i.i.d..
		Thus, we reduce the dimension of the channel state in this Markov chain, which means that the Markov state is the system-AoI-vector.
		To simplify writing, we abbreviate the transition probability as 
		\vspace{-3mm}
		\begin{equation}
			P_{\mathcal{T}_2|\mathcal{T}_1} = \mbox{Pr}\{ \mathcal{T}[n]=\mathcal{T}_2|\mathcal{T}[n-1]=\mathcal{T}_1\}
			\vspace{-3mm}
		\end{equation}
		
		If the transmitter keeps silent in one time slot, then the receiver-AoI would increase by one.
		If the transmission happens, the updated receiver-AoI depends on the transmission rate at this time slot and the buffer-AoI-vector.
		To give the state transition probability of the Markov chain, we present the following theorem.
		Vectors $\mathds{1}_N$ and $\mathds{O}_N$ are $N$-dimensional row vectors, all of whose entries are one and zero, respectively.
		
		\begin{theorem}\label{theorem_1}
			The transition probability of the Markov chain is given by the follow three cases.
			\vspace{-4mm}
			\begin{case}\label{case_1_th}
				When there are no packets in the buffer after the arrival process of the $(n-1)$th time slot, i.e., $\mathcal{T}_1 = (-\mathds{1}_S, A_r)$, the transition probability is given by
				\begin{equation}\label{eq_case_1_th}
				P_{\mathcal{T}_2|\mathcal{T}_1} = 
				\left\lbrace 
				\begin{aligned}
				&\lambda, &&\quad\mathcal{T}_2 = (-\mathds{1}_{S-1}, 0, A_r+1),\\
				&1-\lambda, &&\quad\mathcal{T}_2 = (-\mathds{1}_S, A_r+1).
				\end{aligned}
				\right.
				\end{equation}
			\end{case}
		
			\begin{case}\label{case_2_th}
				When the buffer is not empty and there are $K$ packets, which is less than $S$, in the buffer after the arrival process of the $(n-1)$th time slot, i.e., $\mathcal{T}_1 = (-\mathds{1}_{S-K}, \bm{A}_K^1,A_r)$, where $\bm{A}_m^n$ is the abbreviation for vector $( A_m,A_{m-1}\cdots,A_n),~m\geq n$. Then the transition probability is given by
				\begin{equation}\label{state_transition_2}
					P_{\mathcal{T}_2|\mathcal{T}_1} = 
					\left\lbrace 
					\begin{aligned}
					&\lambda f_{\mathcal{T}_1}^{s}, &&s=0,&&\mathcal{T}_2 = (-\mathds{1}_{S-K-1},0, \bm{A}_K^1+1,A_r+1),\\
					&(1-\lambda) f_{\mathcal{T}_1}^{s}, &&s=0,&&\mathcal{T}_2 = (-\mathds{1}_{S-K}, \bm{A}_K^1+1,A_r+1),\\
					&\lambda f_{\mathcal{T}_1}^{s}, &&s<K,&&\mathcal{T}_2 = (-\mathds{1}_{S-K-1},0, \bm{A}_K^{s+1}+1,A_r+1),\\
					&(1-\lambda) f_{\mathcal{T}_1}^{s}, &&s<K,&&\mathcal{T}_2 = (-\mathds{1}_{S-K}, \bm{A}_K^{s+1}+1,A_r+1),\\
					&\lambda f_{\mathcal{T}_1}^{s}, &&s=K,&&\mathcal{T}_2 = (-\mathds{1}_{S-1},0, A_K+1),\\
					&(1-\lambda) f_{\mathcal{T}_1}^{s}, &&s=K,&&\mathcal{T}_2 = (-\mathds{1}_S, A_K+1),
					\end{aligned}
					\right.
				\end{equation}
			\end{case}
		\vspace{-4mm}
		
			\begin{case}\label{case_3_th}
				When the buffer is not empty and there are $K$ packets, which is no less than $S$, in the buffer after the arrival process of the $(n-1)$th time slot, i.e., $\mathcal{T}_1 = (\bm{A}_S^1, A_r)$, we denote the state of the next time slot by $\mathcal{T}_2 = (\bm{Y_s^1}, \bm{A}_S^{s+1}+1, A_s+1)$. Then the transition probability is given by
				\begin{equation}\label{state_transition_3}
					P_{\mathcal{T}_2|\mathcal{T}_1} = 
					\left\lbrace 
					\begin{aligned}
					&f_{\mathcal{T}_1}^{s}, &&s=0, &&\mathcal{T}_2 = (\bm{A}_S^{1}+1, A_r+1),\\
					&(1-\lambda)^{A_S} f_{\mathcal{T}_1}^{s}, &&s>0,\bm{Y_s^1} =-\mathds{1}_S,&&\mathcal{T}_2 = (-\mathds{1}_S, \bm{A}_S^{s+1}+1, A_s+1),\\
					&\lambda^{\iota}(1-\lambda)^{A_S-\iota} f_{\mathcal{T}_1}^{s}, &&s>\iota,\bm{Y_s^1} \neq-\mathds{1}_S, &&\mathcal{T}_2 = (\bm{Y_s^1}, \bm{A}_S^{s+1}+1, A_s+1),\\
					&\lambda^{\iota}(1-\lambda)^{A_S-s-y_{\iota}} f_{\mathcal{T}_1}^{s}, &&s=\iota,\bm{Y_s^1} \neq-\mathds{1}_S, &&\mathcal{T}_2 = (\bm{Y_s^1}, \bm{A}_S^{s+1}+1, A_s+1).
					\end{aligned}
					\right.
				\end{equation}
			\end{case}
		\end{theorem}
	
		\begin{IEEEproof}
			In Case \ref{case_1_th}, no packet would be transmitted at the $(n-1)$th time slot.
			As the transmitter keeps silent, the state of the $n$th time slot only depends on the arrival process of the $n$th time slot.
			Based on this, we can obtain the transition probability under Case \ref{case_1_th} as shown in Eq. (\ref{eq_case_1_th}).
			
			In Case \ref{case_2_th}, at most $K$ packets can be transmitted at the $(n-1)$th time slot. 
			The state of the $n$th time slot depends on the transmission rate of the $(n-1)$th time slot and the arrival process of the $n$th time slot.
			Based on this, we can obtain the state transition probability under Case \ref{case_2_th} as shown in Eq. (\ref{state_transition_2}).
			
			In Case \ref{case_3_th}, we know that the age of oldest $S$ packets are listed in the $\mathcal{T}_1$ while the other $K-S$ packets are not, which makes the age of the newest $K-S$ packets in the buffer uncertain. 
			Let us denote by $\bm{X_{K-S}^1} = (X_{K-S}, X_{K-S-1},\cdots, X_1)$ the age of the packets whose age is not included in $\mathcal{T}_1$.
			As we do not know the value of the components of $\bm{X_{K-S}^1}$, we characterize $\bm{X_{K-S}^1}$ from a probabilistic way.
			Assume that $s$ packets are transmitted in the $(n-1)$th time slot.
			
%
			
			As the arrival process is independent from the transmission and the scheduling method, the probability distribution of $\bm{X_{K-S}^1}$ is only related to the age of the $S$th oldest packet in the buffer.
			After the $S$th oldest packets have arrived at the buffer, $S$ time slots have passed.
			We also have $K-S\leq A_S$ because at most one packet arrives per time slot.
			Given that the packets arrive with an arrival rate $\lambda$ every time slot, we can obtain the probability distribution of $K$, which follows a Binomial distribution.
			\begin{equation}
			\mbox{Pr}\{K = S + m | A_S = n\} = \binom{n}{m}\frac{1}{2^n},
			\end{equation}
			where $m = 0, 1, \cdots, n$.
			
			As the arrival process across different time slots follow an $i.i.d.$ Bernoulli distribution, the possible values of $\bm{X_{K-S}^1}$ occurs with the same probability.
			\begin{equation}
			\mbox{Pr}\left\{\bm{X_{K-S}^1} = (x_m,x_{m-1},\cdots, x_1)|K = S+m, A_S = n\right\} = \frac{1}{2^n},
			\end{equation}
			where $(x_m,x_{m-1},\cdots, x_1)$  is a possible value of $\bm{X_{K-S}^1}$ and $0\leq x_m<x_{m-1}<\cdots<x_1<n$.
			
			In the $n$th time slot, we denote by $\mathcal{T}_2 = (\bm{Y_s^1}, \bm{A}_S^{s+1}+1, A_s+1)$ the state of the Markov chain, where $s$ is the number of packets transmitted in the $(n-1)$th time slot and $\bm{Y_s^1} = (y_s,y_{s-1},\cdots,y_1)$.
			The state transition probability differs based on the value of $\bm{Y_s^1}$, whose components satisfy $-1\leq y_s\leq\cdots y_1< A_S$.
			If the elements of $\bm{Y_s^1}$ satisfy $\bm{Y_s^1}\neq -\mathds{1}_s$, we assume that the smallest non-negative element in $\bm{Y_s^1}$ is $y_{\iota}$, where $1\leq\iota\leq s$.
			
			When $s=0$, we have		
			\begin{equation}\label{trans_eq_3}
			P_{(\bm{A}_S^{1}+1, A_r+1)|(\bm{A}_S^1, A_r)} = f_{\mathcal{T}_1}^{0}.
			\end{equation}
			
			When $s>0$ and $\bm{Y_s^1} =-\mathds{1}_S$, we have
			\begin{equation}
			P_{(-\mathds{1}_S, \bm{A}_S^{s+1}+1, A_s+1)|(\bm{A}_S^1, A_r)} = (1-\lambda)^{A_S} f_{\mathcal{T}_1}^{s}.
			\end{equation}
			
			When $s>0$ and $\bm{Y_s^1} \neq-\mathds{1}_S$, we have
			\begin{equation}\label{end_eq}
			P_{(\bm{Y_s^1}, \bm{A}_S^{s+1}+1, A_s+1)|(\bm{A}_S^1, A_r)} = 
			\left\{
			\begin{aligned}
			&\lambda^{\iota}(1-\lambda)^{A_S-\iota} f_{\mathcal{T}_1}^{s}, &&\quad \iota<s,\\
			&\lambda^{\iota}(1-\lambda)^{A_S-s-y_{\iota}} f_{\mathcal{T}_1}^{s},&&\quad \iota=s.
			\end{aligned}
			\right.
			\end{equation}
			
			Collectively, we can obtain the state transition probability under Case \ref{case_3_th} as shown in Eq. (\ref{state_transition_3})
		\end{IEEEproof}

		\begin{figure}[tbp]
			\centering
			\includegraphics[width = 0.7\textwidth]{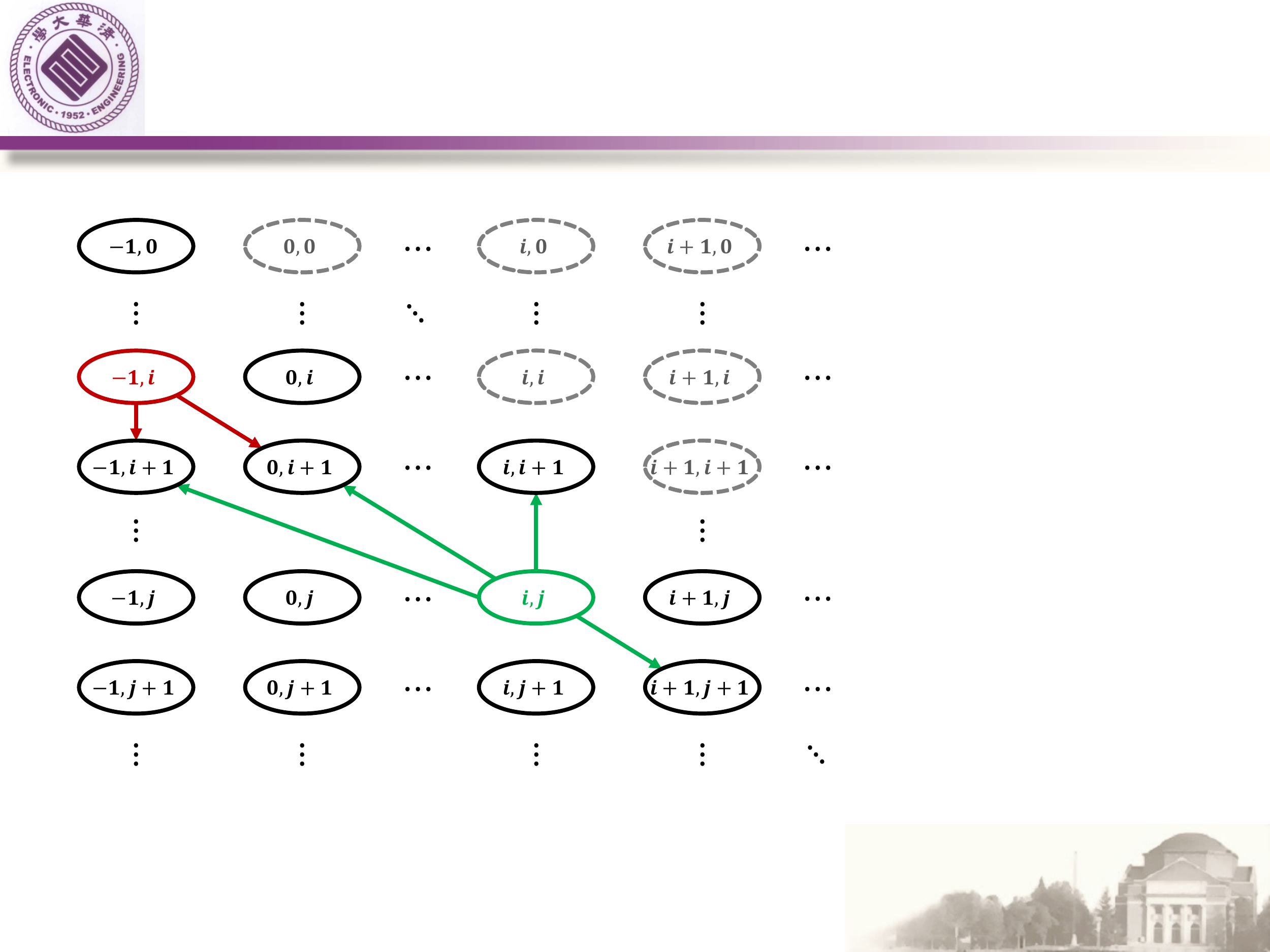}
			\caption{Trellis Graph that Characterize the Legitimate Transitions of System-AoI-vector. In this trellis graph, we set $S = 1$.}\label{state_transit}
		\end{figure}

		In Fig. \ref{state_transit}, the Markov chain model is illustrated when the maximum transmission rate $S=1$. The states with a dot circle are transient states and the states with firm circle are recurrent states.
		Let us denote by $\bm{\Pi}$ the state transition matrix, $\bm{\pi}$ the steady-state probability, and $\mathbb{S}$ the state space of the formulated Markov chain.
		
		We classify the state of the Markov chain into different classes based on the receiver-AoI and the maximum transmission rate. Let us denote the states with the same receiver-AoI $A_r$ and the same maximum transmission rate $S$ by an ordered set $\mathbb{S}_{A_r,S}$. We give the mathematical expression of $\mathbb{S}_{A_r,S}$ through Algorithm \ref{algorithm_state_order}.
		
		\begin{algorithm}[t]
			\caption{Algorithm to Obtain Ordered System-AoI-vector}\label{algorithm_state_order}
			\begin{algorithmic}[1]
				\Require $A_r$, $S$
				\Ensure $\mathbb{S}_{A_r,S}$
				\If {$S=1$}
				\State $\mathbb{S}_{A_r,S} = \left\{(-1,A_r),(0,A_r),\cdots, (A_r-1,A_r)\right\}$
				\ElsIf {$S>1$ and $A_r=0$} 
				\State $\mathbb{S}_{A_r,S} = \left\{(-\mathds{1}_S,0)\right\}$
				\Else
				\State $\mathbb{S}_{A_r-1,S}\leftarrow$ execute Algorithm \ref{algorithm_state_order} with input $A_r-1$ and $S$
				\State $\mathbb{S}_{A_r-1,S-1}\leftarrow$ execute Algorithm \ref{algorithm_state_order} with input $A_r-1$ and $S-1$
				\State $\mathbb{S}_{A_r,S} = \left\{\mathcal{T}|\mathcal{T} = \mathcal{T}'+(\mathds{O}_S,1),\mathcal{T}'\in\mathbb{S}_{A_r-1,S}\right\}\cup\left\{\mathcal{T}|\mathcal{T} = (\mathcal{T}',A_r),\mathcal{T}'\in\mathbb{S}_{A_r-1,S-1}\right\}$
				\EndIf 
				\State \textbf{end algorithm}
			\end{algorithmic}
		\end{algorithm}
		
%
%
		
		For the formulated Markov chain with maximum transmission rate $S$, we split the steady state probability based on the ordered set $\mathbb{S}_{A_r,S}$. We rewrite the steady state probability as
		\begin{equation}\label{steady_state}
		\bm{\pi} = \left(\bm{\pi}_0, \bm{\pi}_1, \bm{\pi}_2,  \cdots\right).
		\end{equation}
		
		The component $\bm{\pi}_i$ is given by
		\begin{equation}
			\bm{\pi}_i = (\pi_1,\pi_2,\cdots,\pi_M),
		\end{equation}
		where $\pi_k$ is the probability of the $k$th state in $\mathbb{S}_{i,S}$ and $M = \left|\mathbb{S}_{i,S}\right|$, where $|A|$ is the number of elements in set $A$.
		
		Based on the segmentation of the state transition probability, we can write the state transition matrix $\bm{\Pi}$ as a partitioned matrix.		
		For two states $\mathcal{T}_i \in\mathbb{S}_{i,S}$ and $\mathcal{T}_j\in\mathbb{S}_{j,S}$, if the transition probability $P_{\mathcal{T}_j|\mathcal{T}_i}$ belongs to the three cases in Theorem \ref{theorem_1}, we can obtain the transition probability from Eqs. (\ref{eq_case_1_th}$-$\ref{state_transition_3}). If not, we have $P_{\mathcal{T}_j|\mathcal{T}_i} = 0$.
		For the sake of discussion, we write the state transition matrix as a partitioned matrix.
		Let us denote the $m$th state in $\mathbb{S}_{i,S}$ by $\mathcal{T}_i(m)$ and the $n$th state in $\mathbb{S}_{j,S}$ by $\mathcal{T}_j(n)$.
		We define a matrix $\bm{\Pi}_{i,j}$, whose value $\bm{\Pi}_{i,j}(m,n)$ at the $m$th row and the $n$th column is given by
		\begin{equation}
			\bm{\Pi}_{i,j}(m,n) = P_{\mathcal{T}_j(n)|\mathcal{T}_i(m)}.
		\end{equation}
		The dimension of matrix $\bm{\Pi}_{i,j}$ is $\left|\mathbb{S}_{i,S}\right|\times\left|\mathbb{S}_{j,S}\right|$.
		
		Then the state transition matrix can be obtained as
		\begin{equation}\label{state_transition_matrix}
		\bm{\Pi}=\left[
		\begin{matrix}
		\bm{\Pi}_{0,0}&\bm{\Pi}_{0,1}&\bm{0}&\bm{0}&\bm{0}&\cdots\\				
		\bm{\Pi}_{1,0}&\bm{\Pi}_{1,1}&\bm{\Pi}_{1,2}&\bm{0}&\bm{0}&\cdots\\
		\bm{\Pi}_{2,0}&\bm{\Pi}_{2,1}&\bm{\Pi}_{2,2}&\bm{\Pi}_{2,3}&\bm{0}&\cdots\\
		\bm{\Pi}_{3,0}&\bm{\Pi}_{3,1}&\bm{\Pi}_{3,2}&\bm{\Pi}_{3,3}&\bm{\Pi}_{3,4}&\cdots\\
		\vdots&\vdots&\vdots&\vdots&\vdots&\ddots\\
		\end{matrix}
		\right],
		\end{equation}
		

		\begin{theorem}
			The state equilibrium equation of the given Markov chain can be obtained as
			\begin{align}\label{end_steady_state}
			\left\{
			\begin{aligned}
			\bm{\pi}_0 &= \sum_{i=0}^{\infty}\bm{\pi}_i\bm{\Pi}_{i,0},\\
			\bm{\pi}_j &= \sum_{i=j-1}^{\infty}\bm{\pi}_i\bm{\Pi}_{i,j}.
			\end{aligned}
			\right.
			\end{align}
		\end{theorem}
	\begin{IEEEproof}
		From the property of Markov chain, we have $\bm{\pi} \bm{\Pi}=\bm{\pi} $ and $\mathds{1}^T \bm{\pi} = 1$. 
		Based on Eqs. (\ref{state_transition_matrix}) and (\ref{steady_state}), we can obtain Eq. (\ref{end_steady_state}).
	\end{IEEEproof}
		
	In summary, the scheduling problem can be formulated into a CMDP which consists of a 4-tuple $\left(\mathbb{S}, \mathbb{A}, \bm{\mbox{Pr}}(\cdot|\cdot), C(\cdot)\right)$, where
	\begin{itemize}
		\item \textbf{System State:} The state of the formulated Markov chain is the system-AoI-vector, which is given in Definition \ref{system_state_def}. The set of all states is denoted by $\mathbb{S}$.
		\item \textbf{Action Set:} At each time slot $n$, there are $S+1$ possible actions $s[n]\in\{0,1,\cdots,S\}$. Therefore, the action set $\mathbb{A}=\{0,1,\cdots,S\}$.
		\item \textbf{Transition Probability:} The state transition probability is given in Theorem \ref{theorem_1}. Moreover, the state transition matrix is given in Eq. (\ref{state_transition_matrix}).
		\item \textbf{Cost Function:} We choose the average power consumption as the cost in our model. For each state $\mathcal{T}$, when $s$ packets are transmitted under channel state $\omega$, the cost function is given by
		\begin{equation}
			C\left(\mathcal{T}\right)=P_{\omega,s}.
		\end{equation}
	\end{itemize}

	Although we have obtained the mathematical expression of the state transition matrix, the analytical expression of the steady state probability remains hard to obtain due to the very large scale of $\bm{\Pi}$.
	Therefore, instead of finding the analytical expression of the steady state probability, we give the expression of AoI and average power consumption and show the inherent relationship between them.
	We give the mathematical expression of AoI and average power consumption through the following theorem.
	
	\begin{theorem}\label{lemma_power_AoI}
		The AoI $A$ and average power consumption $P$ are given by
		\begin{align}
		A &= \sum_{i = 0}^{\infty}\sum_{\mathcal{T}\in\mathbb{S}_{i,S}}i\pi_\mathcal{T}\label{AoI}\\
		P &= \sum_{\mathcal{T}\in\mathbb{S}}\sum_{s = 1}^{S}\sum_{\omega = 1}^{W}\alpha_\omega \pi_\mathcal{T}f_\mathcal{T}^{\omega,s}P_{\omega,s}\label{Power}
		\end{align}
	\end{theorem}
	
	\begin{IEEEproof}
		From Def. (\ref{def_AoI}), we know that the AoI equals the expectation of receiver-AoI $A_r$, i.e., $A = \mathbb{E}\left\{A_r\right\}$.
		The states with the same receiver-AoI $A_r$ are included in vector $\bm{\pi}_i$.
		Therefore, we have
		\begin{equation}\label{temp_1}
		\mbox{Pr}\left\{\mathcal{T} = \left(\bm{A}_S^1, i\right)\right\} = \mathds{1}^T\bm{\pi}_i
		\end{equation}
		
		Based on Eq. (\ref{temp_1}), we have
		\begin{equation}
		\begin{aligned}
		A &= \mathbb{E}\left\{A_r\right\}\\
		& =	\sum_{i=0}^{\infty}i\mbox{Pr}\left\{\mathcal{T} = \left(\bm{A}_S^1, i\right)\right\}\\
		& = \sum_{i=0}^{\infty}i\mathds{1}^T\bm{\pi}_i
		\end{aligned}
		\end{equation}
		
		Till now, we haven proved Eq. (\ref{AoI}).
		For any state $\mathcal{T}\in\mathbb{S}$, the probability that the transmitter transmits $s$ packets under channel state $\omega$ is $\alpha_\omega f_{\mathcal{T}}^{\omega,s}$ and the power consumed is $P_{\omega,s}$.
		Therefore, we have
		\begin{equation}
		P = \sum_{\mathcal{T}\in\mathbb{S}}\sum_{s = 1}^{S}\sum_{\omega = 1}^{W}\alpha_\omega\pi_\mathcal{T} f_\mathcal{T}^{\omega,s}P_{\omega,s}
		\end{equation}
	\end{IEEEproof}
	
\section{Optimal Tradeoff between AoI and Average Power Consumption}\label{analysis}
	In this section, we first formulate an optimization problem to optimize AoI with a given average power constraint. 
	Then, we prove that the optimal scheduling policy can be found in semi-threshold policies. 
	Finally, an algorithm is presented to obtain the optimal scheduling policy.
	
	\subsection{AoI and Average Power Consumption Analysis}	

		From Theorem \ref{lemma_power_AoI}, we know that the AoI and the average power consumption are both determined by the steady state probability $\bm{\pi}$ and the scheduling policy $\bm{F}$. Based on this, we formulate an optimization problem to characterize the optimal tradeoff between AoI and average power consumption.
		\begin{theorem}
			For a given average power constraint $P_c$, the optimal AoI $A^*$ is the solution to 
		\begin{subequations}\label{OP}
			\renewcommand{\theequation}{\theparentequation.\alph{equation}}
			\begin{align}
			\min_{f_{\mathcal{T}}^{\omega,s}}\quad&\sum_{i = 0}^{\infty}\sum_{\mathcal{T}\in\mathbb{S}_{i,S}}i\pi_\mathcal{T},\label{OP_AoI}\\
			s.t.\quad
			&\sum_{\mathcal{T}\in\mathbb{S}}\sum_{s = 1}^{S}\sum_{\omega = 1}^{W}\alpha_\omega\pi_\mathcal{T} f_\mathcal{T}^{\omega,s}P_{\omega,s}\leq P_c,\label{power_constraint}\\
			&\sum_{s=0}^{S}f_{\mathcal{T}}^{\omega,s} = 1,\label{normalization}\\
			&f_{\mathcal{T}}^{\omega,s}\in[0,1],~\mathcal{T}\in\mathbb{S},~1\leq\omega\leq W,~0\leq s\leq S.\label{range}
			\end{align}
		\end{subequations}
	\end{theorem}
		
		\begin{IEEEproof}
		In optimization problem (\ref{OP}), our objective is to minimize AoI in subject to the following constraints: Eq. (\ref{power_constraint}) is the power constraint;
		Eqs. (\ref{normalization}) and (\ref{range}) are the constraints for the scheduling parameters. 
		\end{IEEEproof}
		
		From Theorem \ref{lemma_power_AoI}, we know that the AoI and average power consumption are both determined by $\bm{\pi}$. 
		However, the mathematical expression of the steady state probability is hard to derive due to the large scale of state transition matrix $\bm{\Pi}$. 
		To make the optimization problem (\ref{OP}) solvable, we transform optimization problem (\ref{OP}) into a linear optimization problem through variable substitution.
		\begin{equation}\label{variable_substitution}
			x_{\mathcal{T}}^{\omega,s} = \pi_{\mathcal{T}}f_{\mathcal{T}}^{\omega,s}
		\end{equation}
		
		Based on Eq. (\ref{variable_substitution}), we present the following theorem.
		\vspace{-2mm}
		\begin{theorem}\label{Theorem_1}
			The optimization problem (\ref{OP}) is equivalent to the following linear optimization problem.
			\begin{subequations}\label{OP2}
				\renewcommand{\theequation}{\theparentequation.\alph{equation}}
				\begin{align}
				\min_{x_{\mathcal{T}}^{\omega,s}}\quad&\sum_{i = 0}^{\infty}\sum_{\mathcal{T}\in\mathbb{S}_{i,S}}i\pi_\mathcal{T},\label{OP2_AoI}\\
				s.t.\quad
				&\sum_{\mathcal{T}\in\mathbb{S}}\sum_{s = 1}^{S}\sum_{\omega = 1}^{W}\alpha_\omega\pi_\mathcal{T} f_\mathcal{T}^{\omega,s}P_{\omega,s}\leq P_c,\\
				&x_{\mathcal{T}}^{\omega,s} = \pi_{\mathcal{T}}f_{\mathcal{T}}^{\omega,s}\\
				&\sum_{s=0}^Sx_{\mathcal{T}}^{\omega,s} = \pi_{\mathcal{T}}\\
				&0\leq x_{\mathcal{T}}^{\omega,s}\leq\pi_{\mathcal{T}}.
				\end{align}
			\end{subequations}
		\end{theorem}
	\begin{IEEEproof}
		The details of the proof are given in Appendix \ref{proof_theorem1}.
	\end{IEEEproof}

\subsection{Semi-Threshold Policy Based Algorithm}\label{algorithm_sec}
	This section introduces how to obtain the optimal scheduling policy.
	From Theorem \ref{Theorem_1}, we obtain a linear programming problem, which is much easier compared to optimization problem (\ref{OP}).
	However, due to the infinite number of variables, the optimal tradeoff between the AoI and the average power consumption remains hard to obtain.
	Therefore, we further narrow the variable space through focusing on the semi-threshold policy.
	
	As $\mathcal{F}$ is the set of all policies, we have
	\begin{equation}
		\mathcal{F}=\big\{(f_{\mathcal{T}}^{\omega,s})_{\mathcal{T},\omega,s}|f_{\mathcal{T}}^{\omega,s}\in[0,1],\mathcal{T}\in\mathbb{S},1\leq\omega\leq W,1\leq s\leq S \big\},
	\end{equation}
	which means that the transmission probability depends on the Markov state $\mathcal{T}$, the channel state $\omega$, and the transmission rate $s$. 
	To narrow the strategy space, we try to search through part of the strategies instead of all the strategies. 
	When the receiver-AoI is sufficiently large and the buffer is not empty, transmission can effectively reduce the AoI of the system. 
	Therefore, we focus on semi-threshold-based policies and prove that the optimal AoI can be reached by semi-threshold-based policies.
	The definition of semi-threshold-based policy is given as follows.
	\begin{definition}
		For a scheduling policy $\bm{F}\in\mathcal{F}$, if there exists a positive integer $M$ such that $f_{\mathcal{T}}^{\omega,s} = 1$ when the receiver-AoI $A_r\geq M$, we define $\bm{F}$ as a semi-threshold-based policy and the minimum $M$ as the policy's order.
	\end{definition}
	
	Let us denote by $\mathcal{F}_M$ the set of semi-threshold-based policies with the same order $M$.
	For any scheduling policy $\bm{F}$, we can truncate it at an integer $M$ and obtain a semi-threshold policy. That is, when the receiver-AoI $A_r < M$, the semi-threshold policy choose to transmit with the same probability with policy $\bm{F}$; when the receiver-AoI $A_r \geq M$ and the buffer is not empty, the semi-threshold policy transmits with probability 1.
	Next, we show that any scheduling policy in $\mathcal{F}$ can be approximated by a semi-threshold-based policy.
	\begin{theorem}\label{asymptotic_analysis}
		For any $\varepsilon>0$ and any $\bm{F}\in\mathcal{F}$, if the AoI and the average power consumption under scheduling policy $\bm{F}$ is convergent, there is a positive integer $N$. Such that for any $M\geq N$, there is an $\bm{F_M}\in\mathcal{F}_M$, satisfying $|A_{\bm{F}}-A_{\bm{F_M}}|<\varepsilon$ and $|P_{\bm{F}}-P_{\bm{F_M}}|<\varepsilon$, where $A_{\bm{F}}$, $A_{\bm{F_M}}$, $P_{\bm{F}}$, and $P_{\bm{F_M}}$ denote the AoI and the average power consumption under scheduling policy $\bm{F}$ and $\bm{F_M}$, respectively.
	\end{theorem}
	\begin{IEEEproof}
		The details of the proof are given in Appendix \ref{proof_theorem2}.
	\end{IEEEproof}

	Based on Theorem \ref{finite_dimension_proof}, if we fix the order of the semi-threshold policy at $M$, we can obtain an LP problem with finite variables, which is solvable through linear programming. 
	Let us denote the optimal AoI-power pairs when the system adopt the semi-threshold scheduling policy with order $M$ by $(A_M,P_M)$.
	We denote the optimal solution of LP problem (\ref{OP2}) by $(A_o,P_o)$. 
	Then, from Theorem \ref{asymptotic_analysis}, we know that when $M\rightarrow \infty$, we have $A_M\rightarrow A_o$ and $P_M\rightarrow P_o$.
	As the solution of a semi-threshold policy is sub-optimal, we need to set an acceptable error $\varepsilon$ when we perform numerical calculations.
	Based on this, we present Algorithm \ref{algorithm} to obtain the optimal scheduling policy within an acceptable error $\varepsilon$.
	As the coherence time of the channel is limited, the maximum transmission rate $S$ would not be too large, which guarantees that LP problem (\ref{OP2}) is solvable.
	The complexity of Algorithm \ref{algorithm} is the same order of magnitude as that of linear programming. Assume that the algorithm stops after $L$ iterations. Then the complexity of Algorithm \ref{algorithm} is given by $\tilde{O}((nnz(\bm{\Pi})+d^2)L)$, where $nnz(\bm{\Pi})$ is the number of nun-zero entries in $\Pi$ and $d$ is the number of variables $x_{\mathcal{T}}^{\omega,s}$ \cite{lee2015efficient}.
	\begin{algorithm}[t]
		\caption{Algorithm to Obtain the Optimal Scheduling Policy}\label{algorithm}
		\begin{algorithmic}[1]
			\Require $P_c$, $\varepsilon$, $\lambda$, $S$, $W$, $\{\alpha_1, \alpha_2,\cdots,\alpha_W\}$, $P_{\omega,s},1\leq\omega\leq W,0\leq s\leq S$
			\Ensure $f_{\mathcal{T}}^{\omega,s*}$, $A^*$
			\State initialize $A^*\leftarrow-2\varepsilon$, $A\leftarrow 0$, $M\leftarrow 1$
			\While {$\left(\left|A - A^*\right|>\varepsilon\right)$}
			\State $M\leftarrow M+1$
			\State $A^*\leftarrow A$
			\State $A\leftarrow$ the optimal $A$ obtained by LP problem (\ref{OP2})
			\EndWhile
			\State $x_{\mathcal{T}}^{\omega,s*}\leftarrow\arg\min_{x_\mathcal{T}^{\omega,s}}A$ obtained by LP problem (\ref{OP2})
			\State $\pi_{\mathcal{T}}^*\leftarrow$ substitute $x_{\mathcal{T}}^{\omega,s*}$ into Eq. \ref{inverse_eq}
			\State $f_{\mathcal{T}}^{\omega,s*}\leftarrow$ substitute $\pi_{\mathcal{T}}^*$ and $x_{\mathcal{T}}^{\omega,s*}$ in Eq. \ref{variable_substitution}
			\State $A^*\leftarrow A$
			\State \textbf{end algorithm}
		\end{algorithmic}
	\end{algorithm}


\section{Numerical Results}\label{numerical_results}
	In this section, numerical results are presented to validate the theoretical analysis and demonstrate the efficiency of the presented algorithm. We consider a practical scenario in sensor-based system with Rayleigh fading channels. 
	
\begin{figure}[t]
		\centering
		\includegraphics[width=0.60\textwidth]{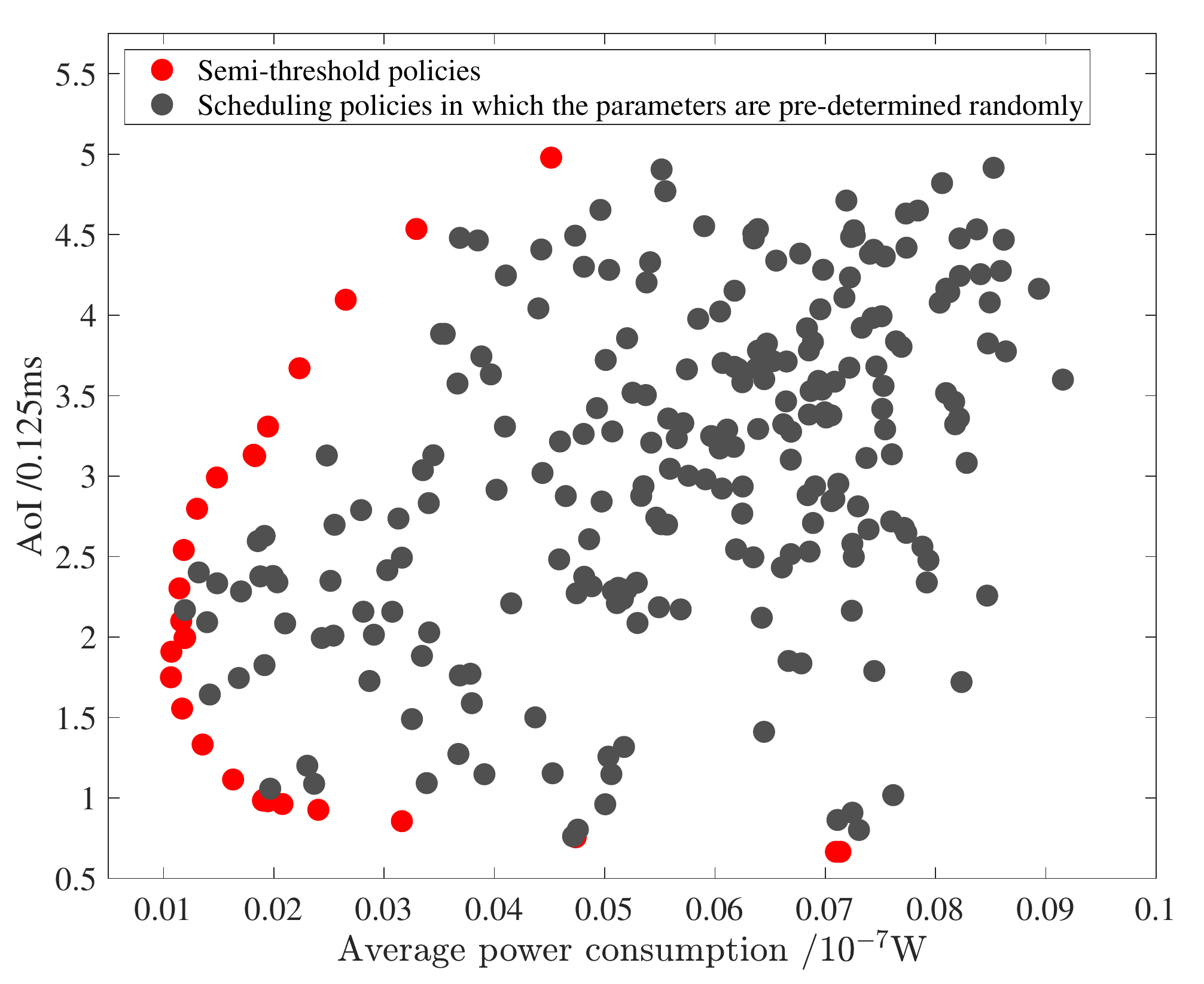}
		\caption{Simulation Results of the AoI-Power Pairs. We compare the proposed semi-threshold policy and the scheduling policy in which the parameters are randomly selected. From the comparison, we see some semi-threshold policies achieve the optimal AoI-power tradeoffs.}\label{semi_threshold}
	\end{figure}

	First, we verify the reduction in complexity of the proposed algorithm. As show in Fig. \ref{semi_threshold}, we fix the arrival rate $\lambda=0.6$, the maximum transmission rate $S=2$, and adopt a Rayleigh fading channel.
	Meanwhile, we set the bandwidth $B=1.5$kHz and the noise power spectral density $N_0=-150$dBm. The length of each time slot is set to 0.125ms and the order of the semi-threshold policy is set to 6. For each deterministic policy, we obtain an AoI-power pair. We use the Monte Carlo method to simulate all the deterministic semi-threshold policies and part of the deterministic non-semi-threshold policies. As shown in Fig. \ref{semi_threshold}, the red dots and the gray dots represent the AoI-power pairs of the deterministic semi-threshold policy and deterministic non-semi-threshold policy, respectively. Because of the large quantity of the deterministic non-semi-threshold policies, we randomly run the simulation of deterministic non-semi-threshold policies for eight times the number of semi-threshold policies. We can see that the semi-threshold policy constitutes the boundary of all the scheduling policies. Therefore, the optimal AoI-power tradeoff can be obtained through searching among semi-threshold policies, which verify the reduction in complexity of the proposed algorithm.
	
\begin{figure}[t]
		\centering
		\includegraphics[width=0.55\textwidth]{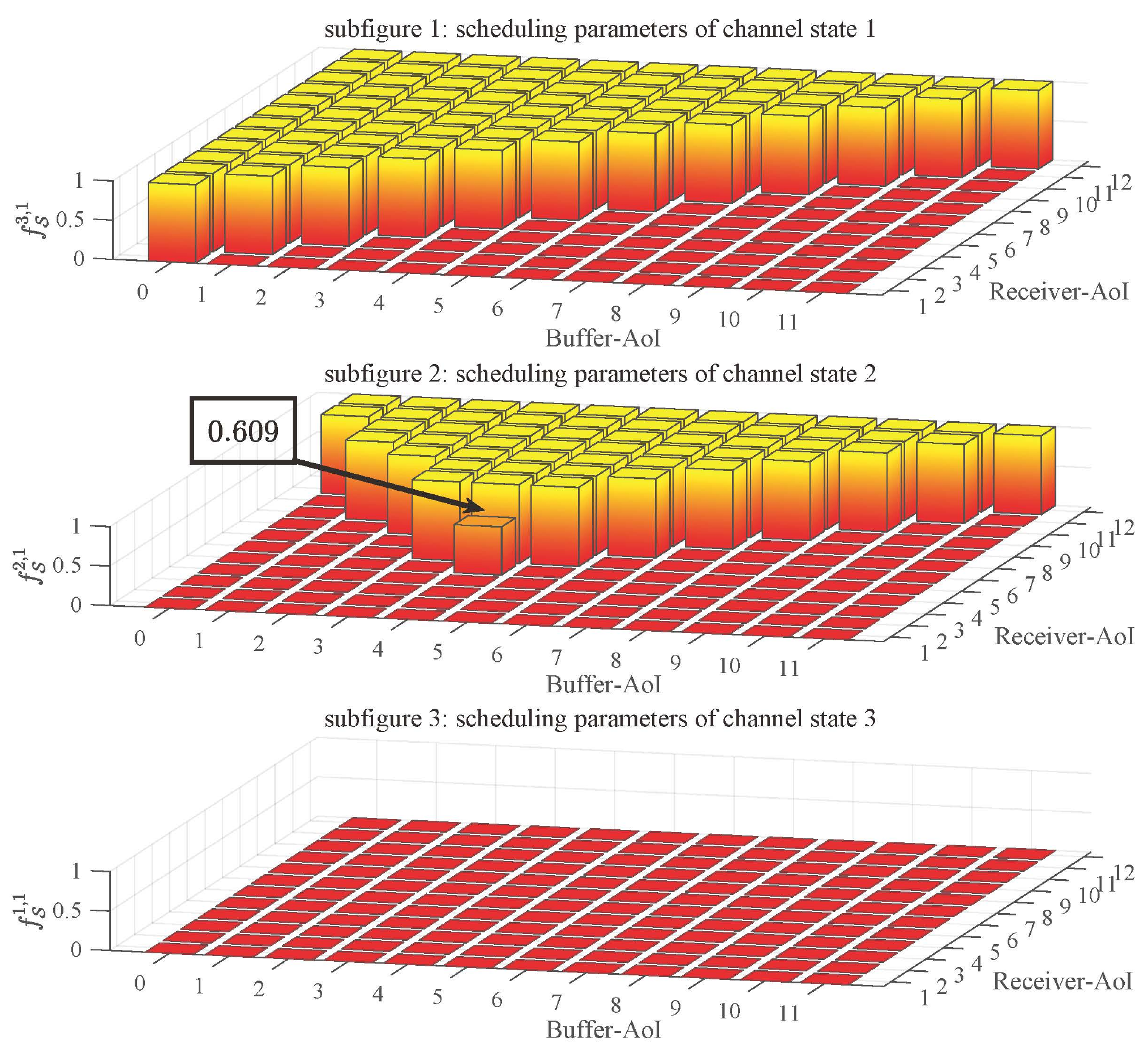}
		\caption{Demonstration of Probabilistic Scheduling Policy: the Probability of Different Transmission Rates}\label{threshold_policy}
	\end{figure}

	Then, we show the threshold structure of the optimal scheduling policy. In this simulation, we fix the arrival rate $\lambda=0.4$, the average power constraint $P_c = 0.848$, and the maximum transmission rate $S=1$.
	We adopt a Rayleigh fading channel and quantize the channel into a three-state channel.
	The bandwidth and the structure of the time slot is the same as the last simulation.	
	By running Algorithm \ref{algorithm} with an acceptable error $\varepsilon=0.1$, we obtain the optimal scheduling policy.
	The algorithm stops when $M=13$ and the obtained policy is shown in Fig. \ref{threshold_policy}.
	Sub-figure 1 shows the scheduling parameters in $\left\{f_{\mathcal{T}}^{3,1}|\mathcal{T}\in\mathbb{S}\right\}$, which corresponds to the best channel condition; sub-figure 2 shows the scheduling parameters in $\left\{f_{\mathcal{T}}^{2,1}|\mathcal{T}\in\mathbb{S}\right\}$, which corresponds to the intermediate channel condition; sub-figure 3 shows the scheduling parameters in $\left\{f_{\mathcal{T}}^{1,1}|\mathcal{T}\in\mathbb{S}\right\}$, which corresponds to the worst channel condition.
	From Fig. \ref{threshold_policy}, we find a distinct threshold structure.
	For the same buffer-AoI-vector and receiver-AoI, the transmitter is more inclined to transmit when the channel state is good.
	Remarkably, the transmitter transmits with probability $0.609$ at state $(4,5)$, which appears to be the threshold of the semi-threshold policy.
	
\begin{figure}[t]
		\centering
		\includegraphics[width=0.60\textwidth]{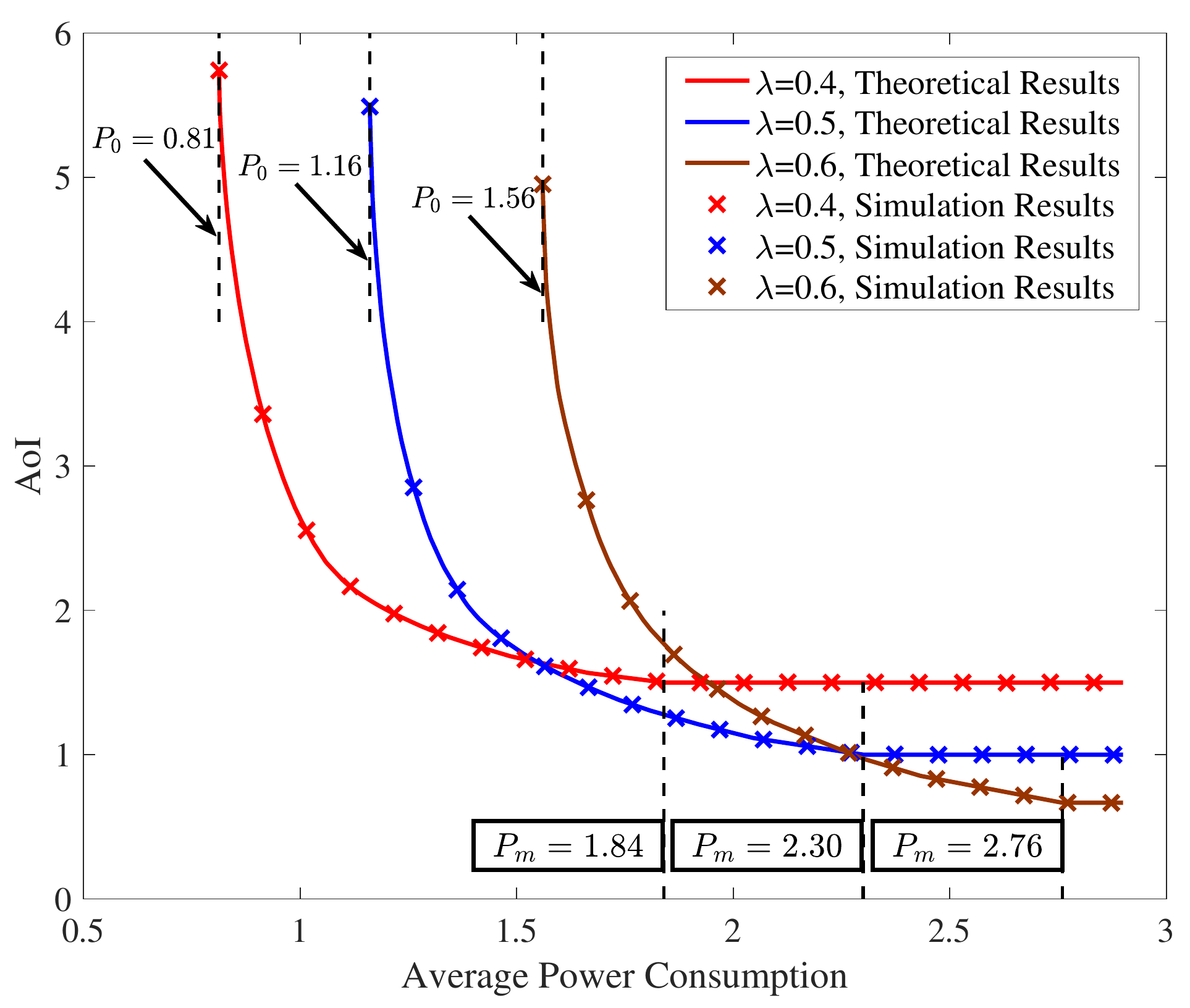}
		\caption{Demonstration of the Optimal AoI-power Tradeoff under Different Arrival Rates}\label{simulation_lambda}
	\end{figure}

	Last but not least, we present the optimal tradeoff between AoI and average power consumption.
	In this simulation, the channel model and time slot structure stay the same. We run the simulation with three different arrival rates, i.e., $\lambda=0.4$, $\lambda=0.5$, and $\lambda=0.6$, respectively.
	By changing the average power constraint form $0.7$ to $2.9$, we obtain the simulation results shown in Fig. \ref{simulation_lambda}.
	The continuous lines in Fig. \ref{simulation_lambda} show the theoretical results of Algorithm \ref{algorithm} while the marker `x' show the Monte Carlo simulation results.
	From the simulation, we find that there is a minimum value for average power consumption to keep the system stable, which is marked as $P_0$.
	There is also an upper bound for average power consumption, which is marked as $P_m$. After the average power consumption exceeds $P_m$, the AoI no longer decreases.
	Moreover, it is worth noting that the optimal arrival rate changes with the power constraint, which indicates that it is necessary to choose a specific arrival rate for different average power constraints.

\section{Conclusions}\label{conclusion}
	In this paper, we have investigated a buffer-aware AoI-optimal scheduling method for wireless transmissions over fading channel.
	By presenting a probabilistic scheduling policy, we have formulated the update process of AoI into a CMDP to minimize the AoI with an average power constraint.
	In the probabilistic scheduling policy, we have taken the buffer-AoI-vector, receiver-AoI, and channel state into account.
	Then, we have converted the tradeoff between AoI and average power consumption into an LP problem.
	Based on the structure of the state transition matrix and steady state probability of the Markov chain, we have further proved that the optimal scheduling policy could be found within the semi-threshold-based policies.
	Further, based on the structure of the optimal policy, a low complexity algorithm has been proposed, with which we can obtain both the optimal AoI-power tradeoff for practical communications.
		
\appendices
\section{The Proof of Theorem \ref{Theorem_1}}\label{proof_theorem1}
	By substituting Eq. (\ref{variable_substitution}) into optimization problem (\ref{OP}), we can obtain the equivalence of optimization problem (\ref{OP}) and optimization problem (\ref{OP2}). Therefore, we just need to prove that optimization problem (\ref{OP2}) is a linear optimization problem of $\bm{x}$.
	First we prove that the steady state probability $\bm{\pi}$ is a linear combination of $\bm{x}$ through the following lemma. 
	
	\begin{lemma}\label{lemma_linear}
		The steady state probability $\bm{\pi}$ can be formulated as a linear function of $\bm{x}$.
	\end{lemma}
	\begin{IEEEproof}
		When the buffer is empty, the transmission would never happen, which means that the transmitter transmits zero packets with probability 1. To make the form of state transition matrix uniform, we rewrite Eq. (\ref{eq_case_1_th}) as
		\begin{equation}\label{eq_case_1_rewrite}
			P_{\mathcal{T}_2|\mathcal{T}_1} = 
			\left\lbrace 
			\begin{aligned}
			& \lambda f_{\mathcal{T}_1}^{0}, &&\quad\mathcal{T}_2 = (-\mathds{1}_{S-1}, 0, A_r+1)\\
			& (1-\lambda) f_{\mathcal{T}_1}^{0},&&\quad\mathcal{T}_2 = (-\mathds{1}_S, A_r+1)
			\end{aligned}
			\right.
		\end{equation}
		where $\mathcal{T}_1 = (-\mathds{1}_S, A_r)$.
		In Eq. (\ref{eq_case_1_rewrite}), we fix $f_{\mathcal{T}_1}^{0} = 1$.
		
		From Theorem \ref{theorem_1}, we notice that every element in the state transition matrix can be expressed in form $C(\mathcal{T}_2,\mathcal{T}_1)f_{\mathcal{T}_1}^s$, where $0\leq s\leq S$ and $C(\mathcal{T}_2,\mathcal{T}_1)$ is given by
		\begin{equation}
			C(\mathcal{T}_2,\mathcal{T}_1) = \frac{P_{\mathcal{T}_2|\mathcal{T}_1}}{f_{\mathcal{T}_1}^s}.
		\end{equation}
		Based on this, the state equilibrium equation of the Markov process can be reformulated as
		\begin{equation}\label{inverse_eq}
			\begin{aligned}
			\bm{\pi}&=\bm{\pi}\bm{\Pi}\\
			&=\bm{x}\bm{C},
			\end{aligned}
		\end{equation}
		where $\bm{C}$ is a matrix composed of $C(\mathcal{T}_2,\mathcal{T}_1)$. Therefore, we know that there is a linear relationship between $\bm{\pi}$ and $\bm{x}$.		
	\end{IEEEproof}

	Based on Lemma \ref{lemma_linear}, the state equilibrium equation and the state normalization equation can both be formulated as a linear restriction of $\bm{x}$.	
	From Eq. (\ref{AoI}), we know that AoI is a linear combination of the elements of $\bm{\pi}$. Thus, the AoI $A$ is also a linear function of $\bm{x}$ because of linear transitivity.
	
	From Eq. (\ref{Power}), we have
	\begin{equation}
		\begin{aligned}
		P &= \sum_{\mathcal{T}\in\mathbb{S}}\sum_{s = 1}^{S}\sum_{k = 1}^{W}\pi_\mathcal{T}\alpha_kf_\mathcal{T}^{\omega,s}P_{\omega,s}\\
		&=\sum_{\mathcal{T}\in\mathbb{S}}\sum_{s = 1}^{S}\sum_{k = 1}^{W}\alpha_kx_\mathcal{T}^{\omega,s}P_{\omega,s}.
		\end{aligned}
	\end{equation}
	Therefore, the average power consumption $P$ is a linear function of $\bm{x}$.
	Moreover, we notice that the scheduling parameters should be in range $[0,1]$. This restriction can be given by $0\leq x_{\mathcal{T}}^{\omega,s}\leq\pi_{\mathcal{T}}$. From Lemma \ref{lemma_linear}, this inequality is also linear.
	Collectively, all the equations in optimization problem (\ref{OP2}) are linear functions of $x_{\mathcal{T}}^{\omega,s}$. Thus, we complete the proof of Theorem \ref{Theorem_1}.

\section{The Proof of Theorem \ref{asymptotic_analysis}}\label{proof_theorem2}
	For a given scheduling policy $\bm{F}$ and any $\varepsilon > 0$, as the AoI is convergent, there exists an integer $N_1$. When $M\geq N_1$, we have
	\begin{align}\label{AoI_limit}
		\sum_{A_r\geq M}A_r\mathds{1}^T\bm{\pi}_{A_r} < \varepsilon.
	\end{align}
	
	We construct a semi-threshold scheduling policy $\bm{F_M}$ that satisfies when $A_r<M$, the scheduling policy $\bm{F_M}$ transmit with the same probability as scheduling policy $\bm{F}$; when $A_r\geq M$ and the buffer is not empty, the scheduling policy $\bm{F_M}$ would transmit with probability 1.
	
	Let us denote by $\bm{\pi}$ and $\bm{\Pi}$ the steady state probability and the state transition matrix of scheduling policy $\bm{F}$. Likely, let us denote by $\bm{\pi}_M$ and $\bm{\Pi}_M$ the steady state probability and the state transition matrix of scheduling policy $\bm{F}_M$. 
	For Further discussion, we first prove that $\bm{\pi}_M$ and $\bm{\Pi}_M$ are equivalent to a finite dimensional Markov process through the following lemma.
	
	\begin{theorem}\label{finite_dimension_proof}
		For a semi-threshold scheduling policy $\bm{F}_M$, its steady state probability $\bm{\pi}_M$ and state transition matrix $\bm{\Pi}_M$ are equivalent to the steady state probability $\bm{\pi}'$ and state transition matrix $\bm{\Pi}'$ of a finite dimensional Markov process.
	\end{theorem}
	\begin{IEEEproof}
		The details of the proof are given in Appendix \ref{proof_finite}.
	\end{IEEEproof}

	Based on Theorem \ref{finite_dimension_proof}, we know that the steady state and state transition matrix of a semi-threshold policy are both finite dimensional matrices. Let us denote the new finite dimensional steady state probability and the state transition matrix by $\bm{\pi}'$ and $\bm{\Pi}'$, respectively. Based on the state equilibrium equation, we have 
	\vspace{-3mm}
	\begin{align}
		&\left\{
		\begin{aligned}
		\bm{\pi}\bm{\Pi}&=\bm{\pi}\\
		\mathds{1}^T\bm{\pi}&=1
		\end{aligned}
		\right.\quad
		\quad\quad\quad
		\left\{
		\begin{aligned}
		\bm{\pi}'\bm{\Pi}'&=\bm{\pi}'\\
		\mathds{1}^T\bm{\pi}'&=1
		\end{aligned}
		\right.\quad
	\end{align}
	\vspace{-3mm}
	
	The steady state probability $\bm{\pi}'$ and the state transition matrix $\bm{\Pi}'$ are both finite dimensional. We extend $\bm{\pi}'$ and $\bm{\Pi}'$ to the same structure as $\bm{\pi}$ and $\bm{\Pi}$ with zeros. That is, for any state $\mathcal{T}$ that is in $\bm{\pi}$ but not in $\bm{\pi}'$, we add the same state to $\bm{\pi}'$ and set its probability at zero. Similar operation is also done to $\bm{\Pi}'$.
	To measure the difference between $\bm{\pi}$ and $\bm{\pi}'$, we have
	\vspace{-3mm}
	\begin{align*}
	\bm{\pi}-\bm{\pi}'  &= \bm{\pi}\bm{\Pi}-\bm{\pi}'\bm{\Pi}'\notag\\
	&= \bm{\pi}(\bm{\Pi}-\bm{\Pi}')+(\bm{\pi}-\bm{\pi}')\bm{\Pi}'.
	\end{align*}
	\vspace{-10mm}
	
	From the definition of the semi-threshold policy $\bm{F_M}$, we know that the state transition matrix $\bm{\Pi}'$ is exactly the same as the corresponding transition probability in $\bm{\Pi}$, i.e., $P_{i,j}=P'_{i,j},~i,~j<M$. Thus we know that the $j$th element in vector $\bm{\pi}(\bm{\Pi}-\bm{\Pi}')$ satisfies
	\vspace{-3mm}
	\begin{align}
	\sum_{i=1}^{\infty}\pi_i(P_{i,j}-P'_{i,j})=\sum_{i=M}^{\infty}\pi_i(P_{i,j}-P'_{i,j}),
	\end{align}
	where $\pi_i$ denotes the probability that the receiver-AoI $A_r=i$.
	
	Noticed that the transition probability belongs to region $[0,1]$, we have
	\vspace{-3mm}
	\begin{equation}
	\begin{aligned}
		\left|\sum_{i=1}^{\infty}\pi_i(P_{i,j}-P'_{i,j})\right|&\leq\sum_{i=M}^{\infty}\left|\pi_i(P_{i,j}-P'_{i,j})\right|\\
		&\leq\sum_{i=M}^{\infty}\pi_i
	\end{aligned}
	\end{equation}
	\vspace{-3mm}
	
	Combined with Eq. (\ref{AoI_limit}), we have
	\vspace{-3mm}
	\begin{equation}\label{limit_value}
		\begin{aligned}
			\left|\sum_{i=1}^{\infty}\pi_i(P_{i,j}-P'_{i,j})\right|&< 
			\frac{1}{M} \sum_{A_r\geq M}A_r\pi_{\mathcal{T}}\\
			&< \frac{\varepsilon}{M}
		\end{aligned}
	\end{equation}
	\vspace{-3mm}
	
	We consider the following simultaneous equations.
	\begin{align}\label{general_equation}
	\left\{
	\begin{aligned}
	\bm{x}\bm{\Pi}'&=\bm{x}\\
	\mathds{1}^T\bm{x}&=0
	\end{aligned}
	\right.
	\end{align}
	
	Based on the generality of scheduling policy $\bm{F}$ and $\bm{F_M}$, we know that the Eqs. (\ref{general_equation}) have unique solutions with probability 1. As $\bm{x}=\bm{0}$ is one solution to Eqs. (\ref{general_equation}), we know that $\bm{x}=\bm{0}$ is the only solution for Eqs. (\ref{general_equation}). When $M\rightarrow\infty$, variable $\bm{\pi}-\bm{\pi}'$ satisfies Eq. (\ref{general_equation}). 
	Till now, we have proved that the steady state probability of scheduling policy $\bm{F_M}$ converge to that of scheduling policy $\bm{F}$ when $M$ approaches to infinity.
	Then, combined with the discussion in Chapter 16 of \cite{CMDP_proof}, we know that for any $\varepsilon >0$, when $M$ is sufficiently large, the AoI and average power consumption of policy $\bm{F}$ and policy $\bm{F}_M$ satisfy $|A_{\bm{F}}-A_{\bm{F_M}}|<\varepsilon$ and $|P_{\bm{F}}-P_{\bm{F_M}}|<\varepsilon$.
	
%
%
%
%

\section{The Proof of Theorem \ref{finite_dimension_proof}}\label{proof_finite}
	For the sake of discussion, we reform the steady state probability as $\bm{\pi}_M = (\bm{\pi}_{-1}, \bm{\pi}_{0+})$, where $\bm{\pi}_{-1}$ denotes the Markov states that the buffer is empty at the present time slot and $\bm{\pi}_{0+}$ denotes the Markov states that the buffer is not empty at the present time slot.
	
	When the buffer is not empty, as our scheduling policy is a semi-threshold policy. The transmitter transmits with probability 1 when the receiver-AoI exceeds $M$. Therefore, the receiver-AoI of the next time slot would definitely decrease when the receiver-AoI of the present time slot is greater that $M$. As our queue method is FCFS, we know that the maximum age of all the packets in the buffer is smaller than $M$. Thus the steady state $\bm{\pi}_{0+}$ is finite dimensional vector.
	
	When the buffer is empty, the transmitter stays silent as there is no packet to transmit. Under this case, the state of the next time slot only depends on the arrival process. If the state at the present time slot is $\mathcal{T}_1=(-\mathds{1}_S,A_r)$, the state $\mathcal{T}_2$ of the next time slot follows
	\vspace{-2mm}
	\begin{equation}
	P_{\mathcal{T}_2|\mathcal{T}_1} = 
	\left\lbrace 
	\begin{aligned}
	&\lambda, &&\quad \mathcal{T}_2 = (-\mathds{1}_{S-1}, 0, A_r+1)\\
	&1-\lambda, &&\quad \mathcal{T}_2 = (-\mathds{1}_S, A_r+1)
	\end{aligned}
	\right.
	\end{equation}
	\vspace{-3mm}
	where $\mathcal{T}_1 = (-\mathds{1}_S, A_r)$.
	
	To character the state transition, we further divide the steady state probability as $\bm{\pi}_M = (\bm{\pi}_{-1}, \bm{\pi}_0,\bm{\pi}_{+})$, where $\bm{\pi}_0=(-\mathds{1}_{S-1}, 0, A_r+1)$. Based on the division of the steady state probability, the state transition matrix can be correspondingly formulated as
	\vspace{-2mm}
	\begin{align}
	\bm{\Pi}_M=\left[
	\begin{matrix}
	\bm{\Pi}_{-1,-1}	&\bm{\Pi}_{-1,0}	&\bm{\Pi}_+\\
	\bm{\Pi}_{0+,-1}	&\bm{\Pi}_{0+,0}	&\bm{\Pi}_{0+,+}
	\end{matrix}
	\right],
	\vspace{-2mm}
	\end{align}
	\vspace{-3mm}
	where the matrix $\bm{\Pi}_{-1,-1}$ and $\bm{\Pi}_{-1,0}$ are given by
	\begin{align}
		\bm{\Pi}_{-1,-1}&=\left[
		\begin{matrix}
		0	&1-\lambda	&0	&\cdots 	&0	&\cdots\\
		0	&0	&1-\lambda	&\cdots 	&0&\cdots\\
		\vdots	&\vdots	&\vdots	&\vdots &\vdots&\cdots\\
		0	&0	&\cdots &0		&1-\lambda&\cdots\\
		\vdots	&\vdots	&\vdots &\vdots		&\vdots&\ddots
		\end{matrix}
		\right],
	\end{align}
	\begin{align}
		\bm{\Pi}_{-1,~0}&=\left[
		\begin{matrix}
		0	&~~~\lambda~~	&0	&\cdots 	&0	&\cdots\\
		0	&0	&~~~\lambda~~	&\cdots 	&0&\cdots\\
		\vdots	&\vdots	&\vdots	&\vdots &\vdots&\cdots\\
		0	&0	&\cdots &0		&~~~\lambda~~&\cdots\\
		\vdots	&\vdots	&\vdots &\vdots		&\vdots&\ddots
		\end{matrix}
		\right].
	\end{align}
	
	If the state at the present time slot is $(-\mathds{1}_S,A_r)$, where $A_r\geq M$, the receiver-AoI of the last time slot should be less than $M$ and the maximum age in the buffer should be less than $M-1$. Thus the state of the last time slot should also belongs to $\bm{\pi}_{-1}$. If the state at the present time slot is $(-\mathds{1}_{S-1},0,A_r)$, similar conclusion can be reached. Therefore, we have
	\begin{equation}
	\pi_{\mathcal{T}} = 
	\left\{
	\begin{aligned}
	&(1-\lambda)\pi_{(-\mathds{1}_S,A_r-1)}, &&\mathcal{T} = (-\mathds{1}_S,A_r),\\
	&\lambda\pi_{(-\mathds{1}_S,A_r-1)}, &&\mathcal{T} = (-\mathds{1}_{S-1},0,A_r),
	\end{aligned}
	\right.
	\end{equation}
	where $A_r\geq M$.
	
	Consider the following two states,
	\begin{equation}
	\pi_{\mathcal{T}}^* = 
	\left\{
	\begin{aligned}
	&\sum_{i=M}^{\infty}\pi_{(-\mathds{1}_S,i)},&&\mathcal{T} = (-\mathds{1}_S,M)\\	
	&\sum_{i=M}^{\infty}\pi_{(-\mathds{1}_{S-1},0,i)}, &&\mathcal{T} = (-\mathds{1}_{S-1},0,M)
	\end{aligned}
	\right.
	\end{equation}
	
	These two state can be regarded as two convergence states. For state $\pi_{(-\mathds{1}_S,M)}^*$, we have
	\begin{equation}
	\begin{aligned}
	\pi_{(-\mathds{1}_S,M)}^* &= \sum_{i=M}^{\infty}\pi_{(-\mathds{1}_S,i)}\\
	&= \sum_{i=M}^{\infty}(1-\lambda)^{i-M}\pi_{(-\mathds{1}_S,M)}\\
	&= \frac{1}{\lambda} \pi_{(-\mathds{1}_S,M)}\\
	&= \frac{1-\lambda}{\lambda} \pi_{(-\mathds{1}_S,M-1)}
	\end{aligned}
	\end{equation}
	
	Then we can rewrite the state equilibrium equation as
	\begin{align}
	\pi_{(-\mathds{1}_S,M)}^* = (1-\lambda)\pi_{(-\mathds{1}_S,M-1)} + (1-\lambda)\pi_{(-\mathds{1}_S,M)}^*
	\end{align}
	
	Similarly, we have
	\begin{align}
	\pi_{(-\mathds{1}_{S-1},0,M)}^* = \lambda\pi_{(-\mathds{1}_{S-1},0,M-1)} + \lambda\pi_{(-\mathds{1}_{S-1},0,M)}^*
	\end{align}
	
	Till now, we know that the matrix $\bm{\Pi}_{-1,-1}$ and $\bm{\Pi}_{-1,0}$  can be replaced by two new matrices $\bm{\Pi}_{-1,-1}^*$ and $\bm{\Pi}_{-1,0}^*$, which are given by
	\begin{align}
	\bm{\Pi}_{-1,-1}^*&=\left[
	\begin{matrix}
	0	&1-\lambda	&0	&\cdots 	&0	\\
	0	&0	&1-\lambda	&\cdots 	&0\\
	\vdots	&\vdots	&\vdots	&\vdots &\vdots\\
	0	&0	&\cdots &0		&1-\lambda\\
	0	&0	&0 &0		&1-\lambda\\
	\end{matrix}
	\right],
	\end{align}
	\begin{align}
	\bm{\Pi}_{-1,0}^*&=\left[
	\begin{matrix}
	0	&~~~\lambda~~	&0	&\cdots 	&0	&\\
	0	&0	&~~~\lambda~~	&\cdots 	&0&\\
	\vdots	&\vdots	&\vdots	&\vdots &\vdots&\\
	0	&0	&\cdots &0		&~~\lambda&\\
	0	&0	&0 &0		&~~\lambda\\
	\end{matrix}
	\right].
	\end{align}
	
	Therefore, when the buffer is empty, the Markov states and the state transition matrix can both be converted to that of a finite dimension Markov process. In summary, we finish the proof of Theorem \ref{finite_dimension_proof}.


\bibliographystyle{IEEEtran}
\begin{spacing}{1.17}
	\bibliography{journal}
\end{spacing}
%

\end{document}